\documentstyle[multicol,aps,epsfig]{revtex}
\input epsf

\newcommand{\be}{\begin{equation}}
\newcommand{\bel}[1]{\begin{equation}\label{#1}}
\newcommand{\ee}{\end{equation}}
\newcommand{\bea}{\begin{eqnarray}}
\newcommand{\ba}{\begin{array}}
\newcommand{\eea}{\end{eqnarray}}
\newcommand{\ea}{\end{array}}

\begin{document}
\draft

%\twocolumn[\hsize\textwidth\columnwidth\hsize\csname@twocolumnfalse%
%\endcsname

\title{Intelligent Controlling Simulation of Traffic Flow in a
Small City Network
 }

\author{M. Ebrahim Fouladvand, M. Reza Shaebani and Zeinab Sadjadi }

\address{Department of Physics, Zanjan University, P.O.
Box 313, Zanjan, Iran.}

\date{\today}

\maketitle

\begin{abstract}

We propose a two dimensional probabilistic cellular automata for
the description of traffic flow in a small city network composed of
two intersections. The traffic in the network is controlled by a
set of traffic lights which can be operated both in fixed-time and
a traffic responsive manner. Vehicular dynamics is simulated and
the total delay experienced by the traffic is evaluated within
specified time intervals. We investigate both decentralized and
centralized traffic responsive schemes and in particular discuss
the implementation of the {\it green-wave} strategy. Our
investigations prove that the network delay strongly depends on
the signalisation strategy. We show that in some traffic
conditions, the application of the green-wave scheme may destructively
lead to the increment of the global delay.

\end{abstract}

%%%%%%%%%%%%%%%%%%%%%%%%%%%%%%%%%%%%%%%%%%%%%%%%%%%%%%%%%%%%%%%%%%%%%%%%
\begin{multicols}{2}
\section{Introduction}

The ever-increasing volume of vehicular traffic flow together with the
lack of money and space in developing further infrastructure of
urban transportation systems are major reasons for seeking
advanced traffic management methods. It has been more than fifty years
since the scientists have intensively tried to understand the
fundamental spatial-temporal aspects of vehicular phenomena to
gain a better insight to the optimal control of vehicular flow.
The phenomena related to the formation of traffic jams have been
attractive to traffic engineers as well as statistical physicists
for many years and now there exists a proliferation of results
both empirically and theoretically in the literature
\cite{may,daganzo,css99,helbing,kerner1,tgf99,tgf01,kerner2}.
The physics contribution to the subject has been rapidly growing since
90'th. Aside from efforts to understand the basic features of a
typical traffic flow via empirical data, over the last decade
physicists have seriously attempted to model the movement of
vehicles to understand fundamental principles governing the
vehicular flow. These investigations have mainly been concerned
with highway traffic. Interested readers can refer to the
elaborate reviews in the literature \cite{css99,helbing}. On the
other hand, notable investigations have, in parallel, dealt with
the challenging subject of city traffic. Physicists' contributions
to city traffic started through the pioneering work of Biham,
Middelton and Levine who developed a two dimensional Cellular
Automaton (CA) model of city network \cite{bml}. The BML model was
later generalized to take into account several realistic features
such as faulty traffic lights, turning of vehicles and green-wave
synchronization etc
\cite{nagatani1,nagatani2,ishibashi,fukui,tadaki1,chung,cuesta,torok,chau,horiguchi,freund,chopard,tadaki2,simon,morako}.
Quite recently, more serious network models, in the framework of
Nagel-Schreckenberg CA model \cite{ns92}, have come into play
which have significantly contributed to the city network
simulation discipline and have opened promising strides in the
city traffic controlling schemes \cite{cs,brockfeld}. Despite
incorporating global signalisation schemes in these very recent
investigations and similar ones proposed by traffic engineers, the
problem of optimisation of traffic flow in a realistic city
network has not yet thoroughly been solved. Diverse degrees of
freedom makes the problem a formidable task. Basically there are
two types of control for traffic lights at city intersections:
{\it fixed-cycle} and {\it traffic responsive}. Fixed cycle
intersections operate with a constant period of time $T$ which,
for each driving direction, is divided into a green period, a
yellow-red period and a red period. Fixed cycle intersections can
be coordinated by {\it offsetting} the start of green period of
consecutive intersections along a desired path in order to create
the so-called green-wave of lights \cite{webster}. In the traffic
responsive scheme, the period of traffic lights, and consequently
its sub-phases, are simultaneously adapted to the traffic
characteristics and are therefore not constant in the course of
time. It is now an almost well-established fact that, in general,
adaptive controlling of coordinated intersections should be the
ultimate method towards the global signalisation of traffic
lights. The complexity of the problem still remains regarding the
fact that there are numerous schemes of adaptive controlling.
Examining new control schemes on real traffic is not always
possible and practical. Alternatively computer-based simulations
offer a cheap and useful method for testing new strategies that
can be of practical relevance for various applications in city
traffic. It is our objective in this paper to simulate the flow of
vehicles at a simplified set of intersections which operate
under adaptive controlling schemes. Our investigations aim at
seeking the optimal control strategy and to gain a deeper insight
into the problem of city traffic control.

\section{ Formulation of the Model}

We begin our investigations by considering the most simplified
city network i.e., a cluster of two intersections. In our
primitive network, two south-north streets A and B are intersected
by a west-east street C. Figure one illustrates the situation. For
simplicity we assume the south-north streets allow
 single-lane traffic flows northward and that the west-east
street conducts an eastward single-lane flow. A set of traffic
lights controls the traffic at each intersection. Each street is
modelled by a chain divided into cells which are as  large as a
typical car length . Each cell can be either occupied by a car or
being empty. The car velocity can take discrete-valued velocities
$1,2,\cdots, v_{max}$. To be more specific, at each step of time,
the system is characterized by the position and velocity
configurations of cars and the traffic light states at each road.

\begin{figure}\label{Fig1} \epsfxsize=7.5truecm
\centerline{\epsfbox{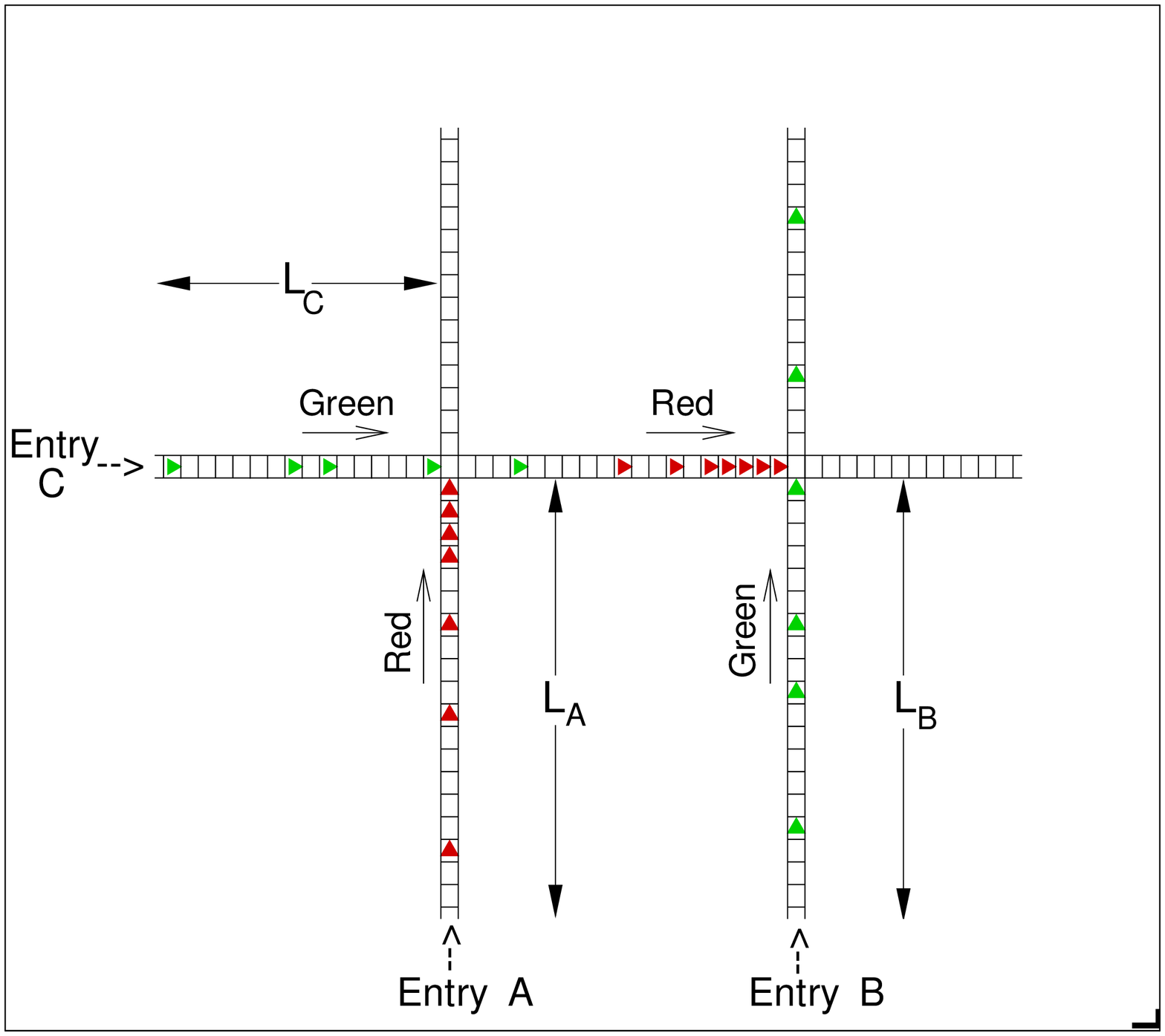}}
\end{figure}
\vspace{0.02 cm} {\small{Fig.~1: Small city network: intersection
of one west-to-east street with two parallel south-to-north
streets. All the streets conduct uni-directional
single-lane traffic flow. }}\\

The movement dynamics is governed by a generalized version of
the Nagel-Schreckenberg model which has recently been proposed by
Knospe, Santen, Schadschneider and Schreckenberg \cite{knospe}.
The model basically differs from the NS model in the sense that it
incorporates adaptation effects, absent in the minimal model of
NS, in realistic traffic. The model successfully reproduces, on a
microscopic level, the generic features of empirical data. The
model incorporates additional parameters and concepts such as
brake lights, safety distance, temporal interaction horizon, finer
discretization of space etc which are the ingredients of an
adaptive look-ahead driving strategy \cite{kerner3}. Let us
briefly explain the movement rules. We denote the position,
velocity and space gap (distance to its leading car) of a typical
car at discrete time $t$ by $x^{(t)},v^{(t)}$ and $g^{(t)}$ and
the same quantities for its leading car by $x_l^{(t)},v_l^{(t)}$
and $g_l^{(t)}$. Assuming that the expected velocity of the
leading car, anticipated by the one following, in the next time step
$t+1$ takes the form $v_{l,anti}^{(t)}=min(g_l^{(t)},v_l^{(t)})$,
we define the effective gap as $g_{eff}^{(t)}:= g^{(t)} +
max(v_{l,anti}^{(t)}-gap_{secure},0)$ in which $gap_{secure}$ is
the minimal security gap. Concerning the above considerations, the
following updating steps evolves the position
and the velocity of each car.\\

1) Acceleration:\\

$v^{(t+1/3)}:= min(v^{(t)}+1, v_{max})$\\

2) Velocity adjustment :\\

$v^{(t+2/3)}:=min(g_{eff}^{(t+1/3)}, v^{(t+1/3)})$\\

3) Random breaking with probability $p$:\\

if random $< p$ then $v^{(t+1)}:=max(v^{(t+2/3)}-1,0)$\\

4) Movement : $x^{(t+1)} :=x^{(t)}+ v^{(t+1)}$ \\

We now return to our network. Let us now specify the physical
values of our time and space units. Ignoring the possibility of
existence of long vehicles such as buses, trucks etc, the length
of each cell is taken to be 5.6 metres which is roughly the
typical bumper-to-bumper distance of cars in a waiting queue.
Concerning the fact that in most urban areas a speed-limit of 60
kilometre/hour should be kept by the drivers, we quantify the time
step in such a way that $v_{max}=6$ corresponds to the speed-limit
value (60 km/h). In this regard, each time step equals two seconds
and correspondingly each discrete increment of velocity signifies
a value of 10 km/h. We set the length of streets before
the intersections (horizon length), denoted by $L_A, L_B$ and $L_C$
respectively, equal to 70 cells and the distance between
intersection as 40 cells . The state of the system at time $t+1$
is updated from that in time $t$ by the synchronous application of the
anticipated NS dynamical rules to all the
vehicles through the following steps:\\

{\bf Step 1 : signal determination}.\\

We first specify the signal states for all of the driving
directions. In subsequent sections we will, in detail, explain the
scheme at which the traffic lights change their colour.\\

{\bf Step 2 : movement at the green roads}.\\

At this stage, we update the position and velocities of cars on
the green road according to the dynamical rules which are
synchronously applied to each car.\\

{\bf Step 3 : movement at the red roads, delay evaluation. }\\

Here the updating is divided into two parts. In the first part, we
evaluate, for each street, the delay of cars waiting on the red
period of the signalisation. In the second half, we update the
position and velocities of the moving cars approaching the waiting
queue(s). We recall that once the light turns red, the
moving cars continue their movements until they come to a complete
stop by reaching to the end of the waiting queue. As soon as a car
comes to halt, it contributes to the total delay. In order to
evaluate the delay, we measure the queue length (the number of
stopped cars) at time step $t$ and denote it by the variable $Q$.
Denoting the configuration of site $i$ ($i$ increases in the
opposite direction of traffic flow) at time $t$ by
$pos^{red}[i,t]$, we have $ pos^{red}[i,t]=1 $ for $i=1 \cdots Q $
and zero at $i=Q+1$. Delay at time step $t+1$ is obtained by
adding the queue length $Q$ to the delay at time step $t$. \be
delay(t+1)= delay(t)+ Q(t) \ee

This ensures that during the next time step, all the stopped cars
contribute one time step to the delay. The next part of the update
goes to the positions and velocities of the moving cars. Moving cars
can potentially be found in the cells $ Q+2,Q+3, \cdots, L$. We
update their positions and velocities in a similar manner
described for green streets.\\

{\bf step 4 : entrance of cars to the intersection}.\\

So far, we have dealt with those cars within the horizons of the
intersections. Here we discuss the entrance of cars into the
network. Evidently from our everyday driving experience, we
observe that the time head-ways between entering cars vary in a
random manner which consequently implies a random distance headway
between successive entering cars. As a candidate for describing
the statistical behaviour of random space gap of entering cars, we
have chosen the Poisson distribution. Very recent empirical studies on
high way traffic confirms that in some traffic phases, the
distribution of the time headways between successive cars agrees
well the Poisson distribution function
 \cite{schad-emp}. According to this distribution function, the probability that the
space gap between the car entering the intersection horizon and
its predecessor be $n$ is : $ p(n) = \frac{\lambda^n e^{-\lambda}
}{n!} $ where the parameter $\lambda$ specifies the average as
well as the variance of the distribution function. The parameter
$\lambda$ is of direct relevance to the traffic volume. A large
value of $\lambda$ describes light traffic while on the other
hand, a small-valued $\lambda$ corresponds to heavy traffic.
Car-injection to the streets has been thoroughly discussed in
\cite{fool-saj-shab}.

The above {\it ad hoc} rules updates the configuration of the
intersections at time $t$. In the course of time the cars enter
the intersections and a fraction of them experience the red lights
and consequently have to wait until they are allowed to go through
intersections during the upcoming green periods. For the sake of
simplicity, we have assumed that each street has a single lane.
For streets with more than one lane, one simply should multiply
the value of delay by the number of lanes. Moreover, we allow
those fractions of northward (eastward) cars tending to turn right
(left) manage to turn via a by-pass road. Therefore the cars in
our simulation denote those which wish to go through the
intersection without turning. We are now able to simulate the
performance of this small network and obtain the delay for each
intersection.

\section{ Signalisation of traffic lights: Traffic Responsive }

We now discuss the main essence of the paper which is the
simulation of intelligent signalisation schemes. Nowadays advanced
traffic control systems anticipate the traffic approaching
intersections. Traffic-responsive methods have shown a very good
performance in controlling the traffic in city network and now a
variety of schemes exists in the literature
 \cite{webster,bang,gartner,cowell,robertson,huberman}. In principle, there
 are two major strategies
for controlling an intersection in the traffic-adaptive methods:
{\it centralized} and {\it decentralized}. In decentralized
schemes, the traffic lights are adapted to the local traffic at
the intersection and are not influenced by the traffic volume
approaching the intersection from adjacent intersections. On the
other hand, in a centralized scheme, the signalisation is
performed in a more intelligent manner and incorporates the
traffic states at the up stream intersections. Whether to control
an intersection via centralized strategy is a subtle question and
often controversial. While empirical studies confirm that in
special circumstances, decentralized local adaptive strategies
operate more effectively than globally adaptive ones
\cite{huberman}, the general trend in controlling central areas of
cities is toward the adaptive coordination of intersections. We
first discuss the decenteralized adaptive controlling scheme.

\subsection{decenteralised adaptive scheme}

 In this scheme, each intersection adapts its signalisation to the local
 traffic in its vicinity and does not take into account the up-stream traffic state in the
 neighboring intersections. The data obtain via traffic detectors installed at
 the intersection is gathered for each movement direction and it is possible to
 measure the queue lengths. One can also
measure the time-headways between successive cars passing each
lane detector. Thus it is possible to estimate the traffic volume
existing at the intersection. There are various methods for
distributing green time to streets. Let us first introduce
$T_{min}$ which is the minimum green time devoted to a direction.
If a typical direction goes green it will definitely remain green
for $T_{min}$ seconds for practical purposes the main one of which
is the slow movement of standing vehicles. Now we try to explain
some standard termination algorithms. In each scheme, the green
time of a typical green street is terminated if some conditions
are fulfilled. By green (red) street we mean the street for which
the traffic light is green (red). We now state the termination
schemes.\\

{\bf Scheme (1)}: The queue length in the conflicting direction
exceeds a cut-off value $L_c$. This scheme only concerns
the traffic states in the red street.\\

{\bf Scheme (2)}: The time headway between successive cars,
denoted by $T_h$, going through the green light exceeds the
cut-off value $T_h^{c}$. Here the algorithm only concerns the
traffic state in the green street.\\

{\bf Scheme (3)} : Each direction is endowed with two control
parameters $L_c$ and $T_h^{c}$. The green phase is terminated if
the conditions: $ T_h ^{g} \geq T_h^{c} $ and $L^{r} \geq L_c$ are
both satisfied.\\

Here the algorithm implements the traffic states in both streets.
The superscripts "r" and "g" refer to words "red" and "green"
respectively. We note that the first two schemes are special cases
of the more general scheme (3). Schemes (1) and (2) are the
limiting behaviour of schemes (3) by letting $T_h^{c} \rightarrow
0$ and $L_c \rightarrow 0$ respectively. In general, the numerical
value of control parameters $L_c$ and $T_h^{c}$ could be taken
different for each individual street. In \cite{fool-saj-shab} we
have shown that scheme (1) is the optimal algorithm for isolated
intersections. In what follows we present our simulation results
for the first signalisation scheme introduced above.

\subsection{ Simulation Results }

We let the network evolve for 1800 time steps which is equal to a
real time period of one hour. We evaluate the aggregate delay for
both intersections. We first consider the
symmetric traffic state in which the traffic conditions are equal
for all streets. In this case, we equally load the intersections
with entering cars spatially separated by random space gap,
obeying the Poisson statistics, from each other. Fig. (2) depicts the total
delay curves for some various cut-off lengths.\\

\begin{figure}\label{Fig2}
\epsfxsize=7.5truecm \centerline{\epsfbox{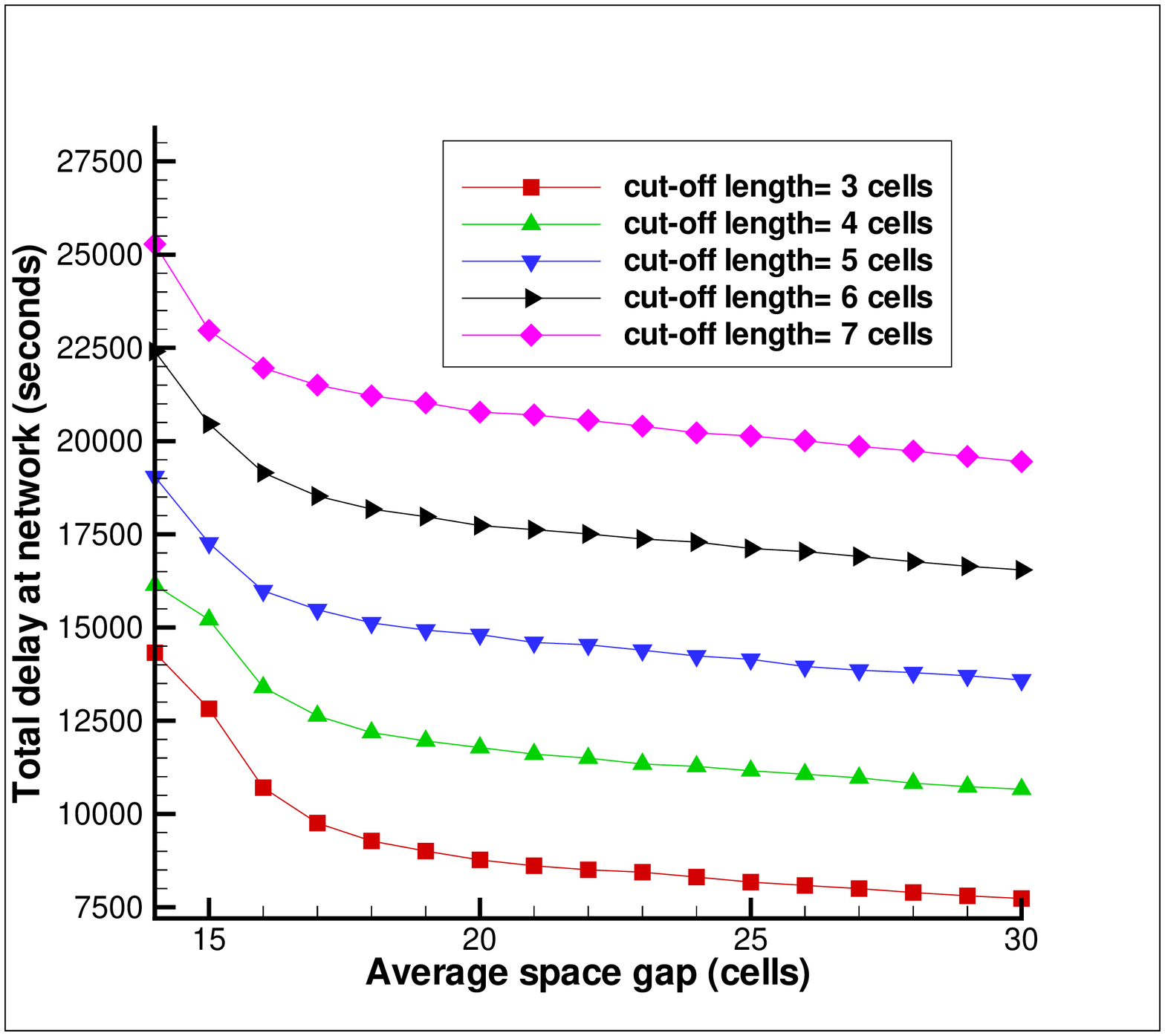}}

\vspace{0.02 cm} {\small{Fig.~2:  Total delay of the network in
terms of mean space gap of entering cars for various cut-off queue
lengths (taken equal for all streets. }}
\end{figure}

We note that by increasing $\lambda$ i.e. decreasing traffic
volume, the delay decreases. For a fixed value of $\lambda$, the
lower cut-off length leads to lower delay. This is expected since
the lower $L_c$ terminates the queue sooner and hence less
contribution is given to the delay. Furthermore, we have examined
the behaviour of delay curves for a wide range of control
parameters space $(L_c^{A},L_c^{B},L_c^{C})$. The following graphs
(figure 5) exhibit this behaviour for fixed traffic volumes. We
have taken $L_c^{A}=L_c^{B}$.

\begin{figure}\label{Fig3}
\epsfxsize=7.5truecm \centerline{\epsfbox{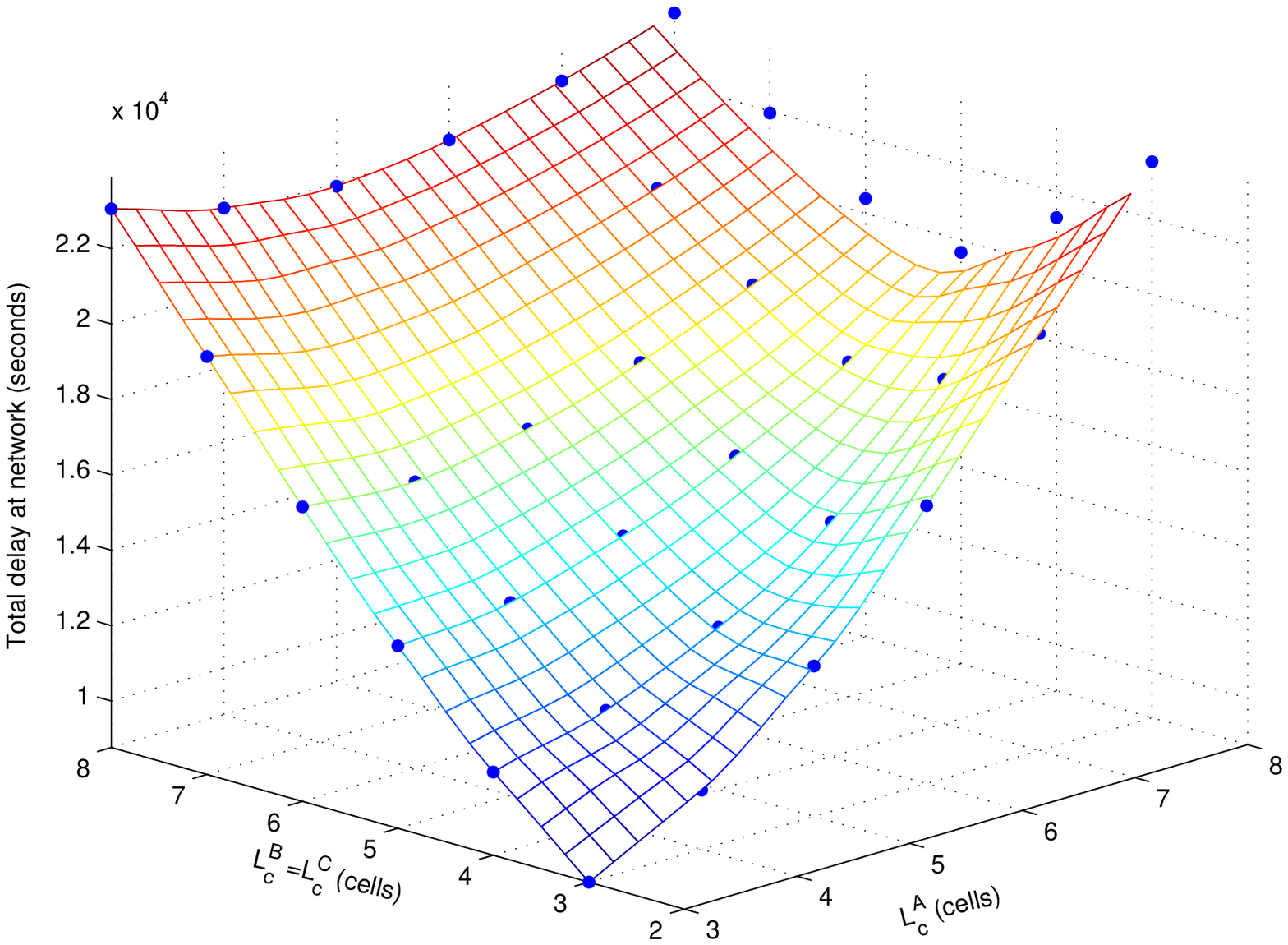}}

\end{figure}

\begin{figure}\label{Fig3}
\epsfxsize=7.5truecm \centerline{\epsfbox{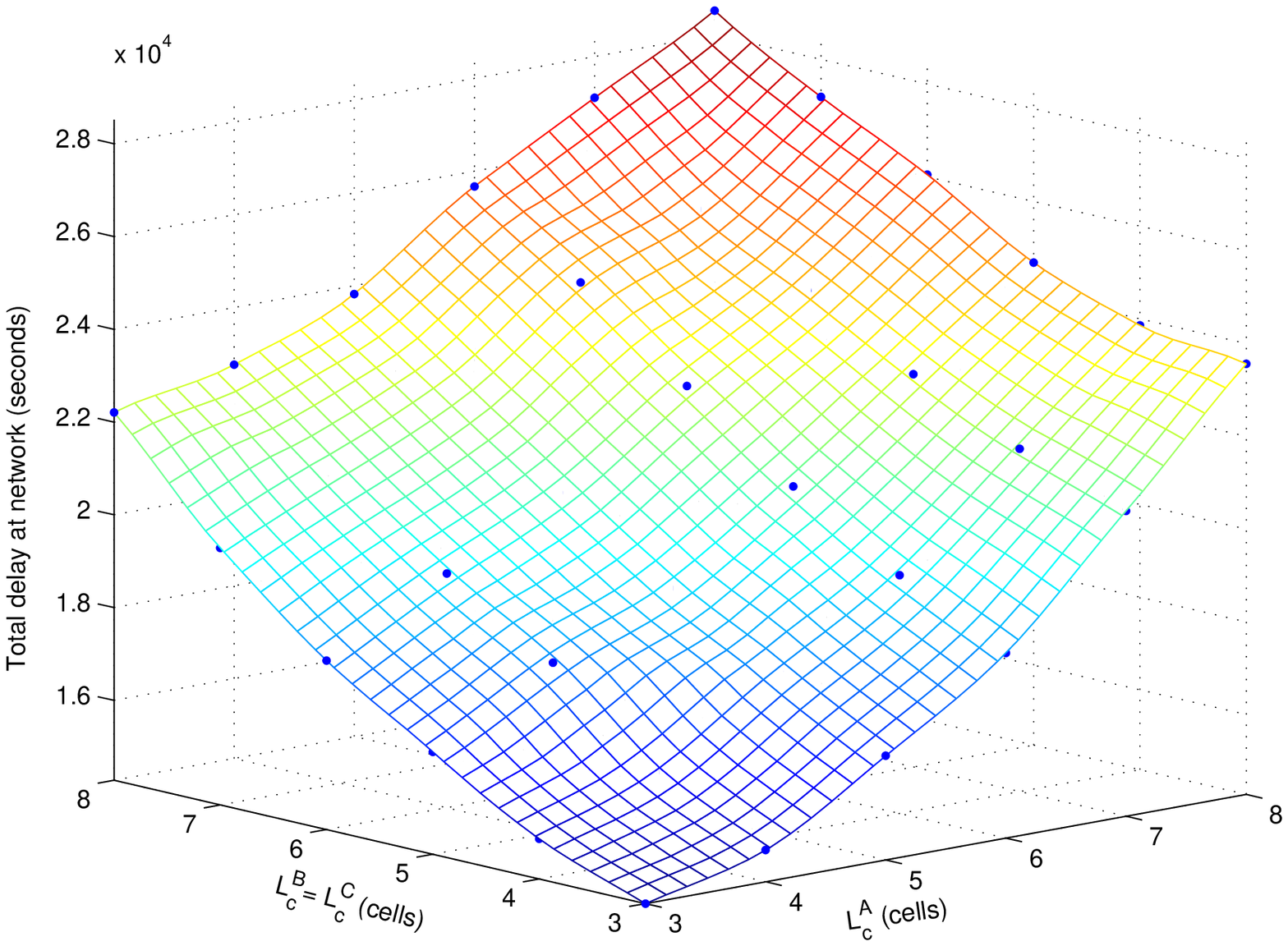}}

\vspace{0.02 cm} {\small{Fig.~3: Overall network delay for a wide
range of the control parameter space. Traffic volume is
symmetrically taken as $\lambda=14$ cells (top graph) and 20
(bottom graph). }}

\end{figure}

The above graphs confirms the conclusion that the optimal
algorithm is the one in which all the cut-off lengths are equal
and taken as short as possible. Let us now investigate the situation
in which two of the streets
are major while the third one is minor. Three cases are
distinguished corresponding to minority of streets A, B and C
respectively. Fig (4) exhibits the results for the case where
street A is minor. The traffic volume at major streets are fixed
at $\lambda=14$ cells.

\begin{figure}\label{Fig4}
\epsfxsize=7.3truecm \centerline{\epsfbox{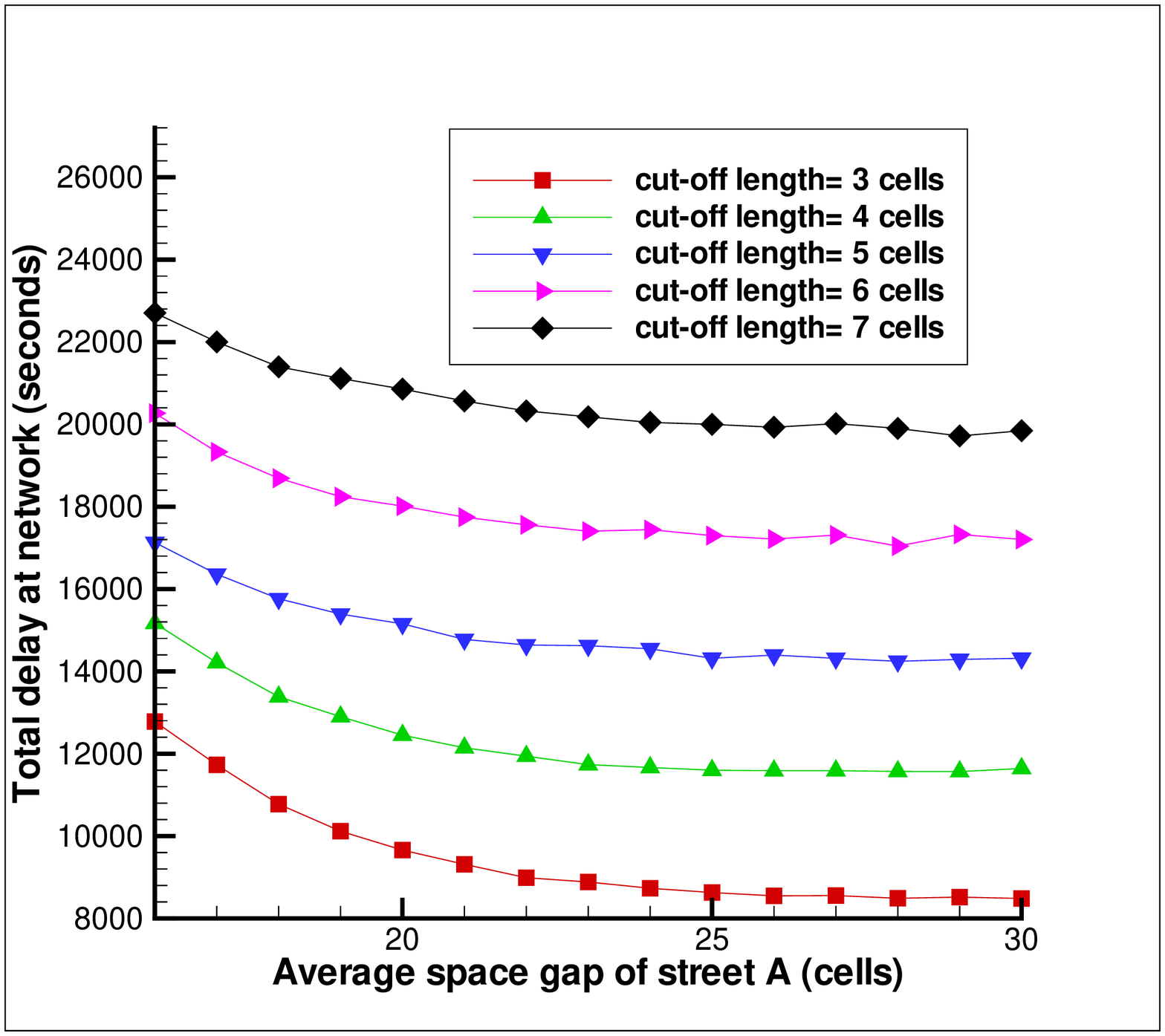}}

\vspace{0.02 cm} {\small{Fig.~4: Overall network delay versus the
traffic volume of the minor street A for various cut-off lengths.
Traffic volume at the major streets are $\lambda=14$ cells. }}

\end{figure}

The result for the cases where the minor street is taken as B an C
are similar to the above graph. It would be useful to analyze the
delay in the individual intersection. In the case where the minor
street is B, preventing the direct intersection of two major
streets A and C, simulation shows the delay at the down-stream
intersection is not a decreasing function of the traffic volume in
the minor street B. Fig (5) depicts the result. For all the volume
range in which there are two major streets and one minor one, the
optimal algorithm is the one with shortest cut-off lengths.

\begin{figure}\label{Fig5}
\epsfxsize=7.2truecm \centerline{\epsfbox{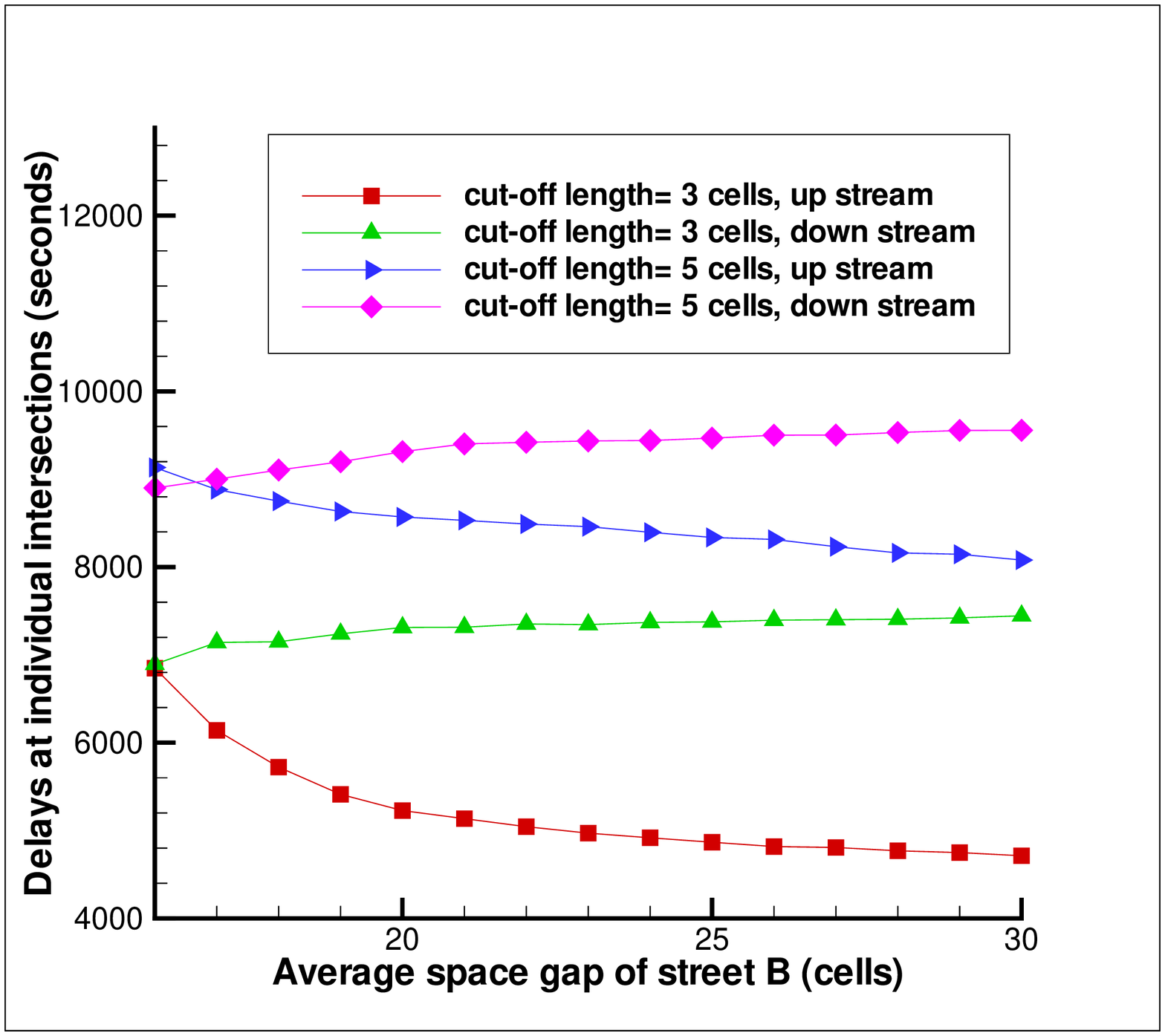}}

\vspace{0.02 cm} {\small{Fig.~5: Overall as well as individual
delays. Cut-off lengths are equal for all streets. Traffic volume
is taken as $\lambda=14$ cells in major streets A and C. }}

\end{figure}

\subsection{centeralized adaptive scheme: green-wave method}

We shall now turn to the main issue of the paper and present our
simulation results of the centralized responsive method. Here the
downstream intersection adapts its signalisation to the
approaching traffic flow from the up-stream one. In this paper we
restrict ourselves to the {\it green-wave} strategy in which the
green time of the street C at intersection with street B is set in
such a way that the platoon coming from the up-stream intersection
can go through the intersection without being interrupted by the
light changing to red. The up-stream intersection, the platoon source,
acts as an independent intersection in a manner explained earlier.
Its signalisation is performed by scheme (1). Note that the
platoon length is determined by the $L_c$ in the up-stream
intersection.

\begin{figure}\label{Fig6}
\epsfxsize=7.2truecm \centerline{\epsfbox{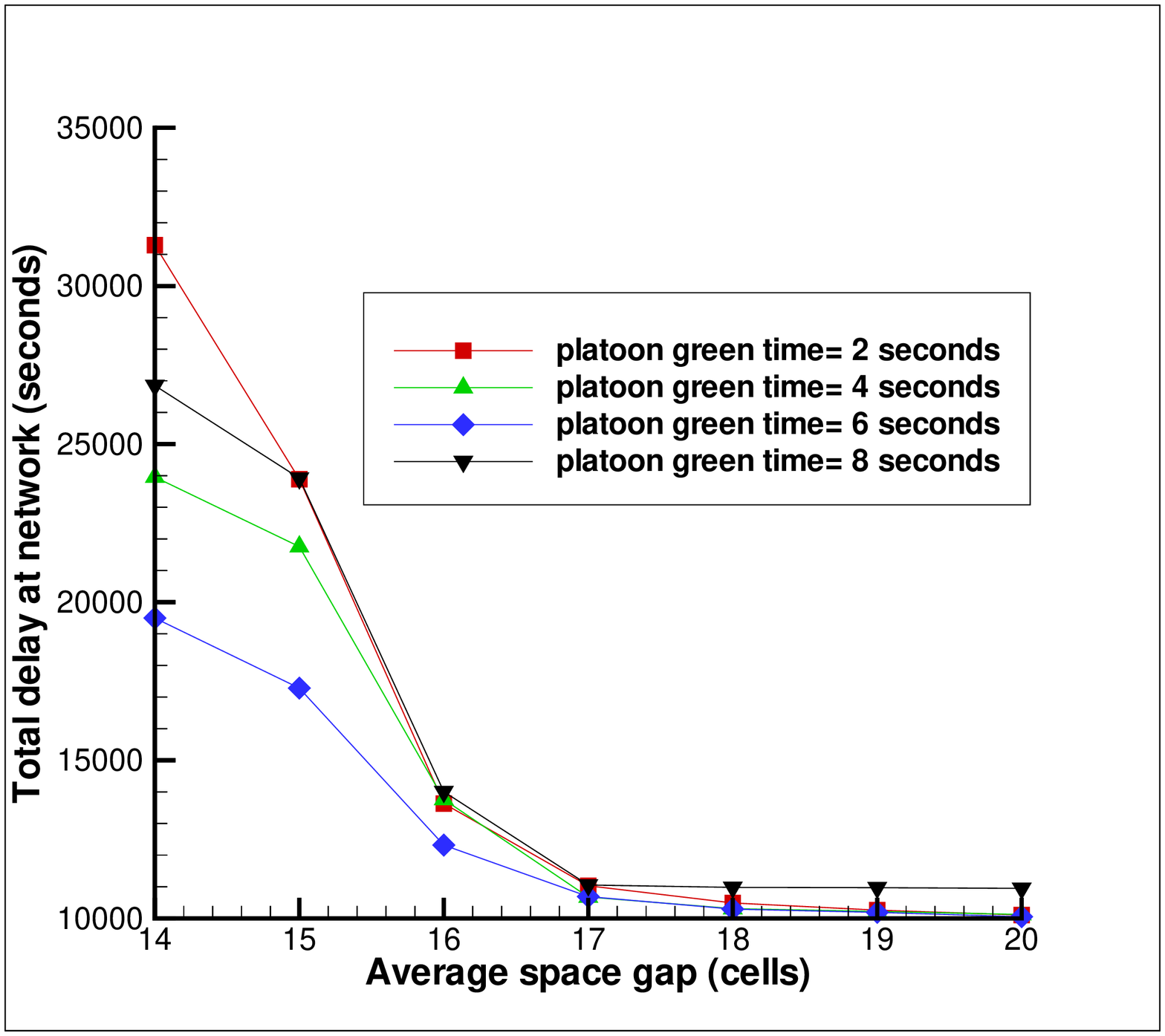}}

\vspace{0.02 cm} {\small{Fig.~6: Overall network delay in terms
of traffic volume (taken equal for all streets) in the green-wave
strategy. Cut-off lengths are equal to five in the up-stream
intersection. }}

\end{figure}

In figure (6) we exhibit the simulated overall delay for various
green time interval given to the approaching platoon. In the time
interval between the passage of the platoon and arrival of the
next platoon, the downstream intersection is controlled by
adaptation to its local traffic. According to the figure, the
optimal timing is obtained by taking the platoon green-time at
$t=6$ seconds. This corresponds to the time required for the
entire passage of a platoon of the length 100 metre moving with
the maximum velocity of 60 km/h. It would be plausible to assume
that the vehicles inside the platoon are separated by an average
headway of car length (4.5 metres). Therefore the average length of
an N-platoon is $5.6N + 4.5(N-1)$. In the above graph, $N$ equal
to cut-off length at the up-stream intersection which is taken as
$N=5$ vehicles. Correspondingly our platoon length is 48 metres
which gives the required time roughly 3 seconds. However, we
should consider that for practical reasons, the lights in the
downstream intersections should go green at least several seconds
before the arrival of the incoming platoon to the border of the
intersection. Otherwise the platoon slows down because of the red
light and this contradict the philosophy of the green-wave method.
In figure (7) we compare the predictions of the green wave method to
that of decenteralized scheme. It exhibits the overall delay in
terms of traffic volume which is assumed to equal for all three
streets.

\begin{figure}\label{Fig7}
\epsfxsize=7.5truecm \centerline{\epsfbox{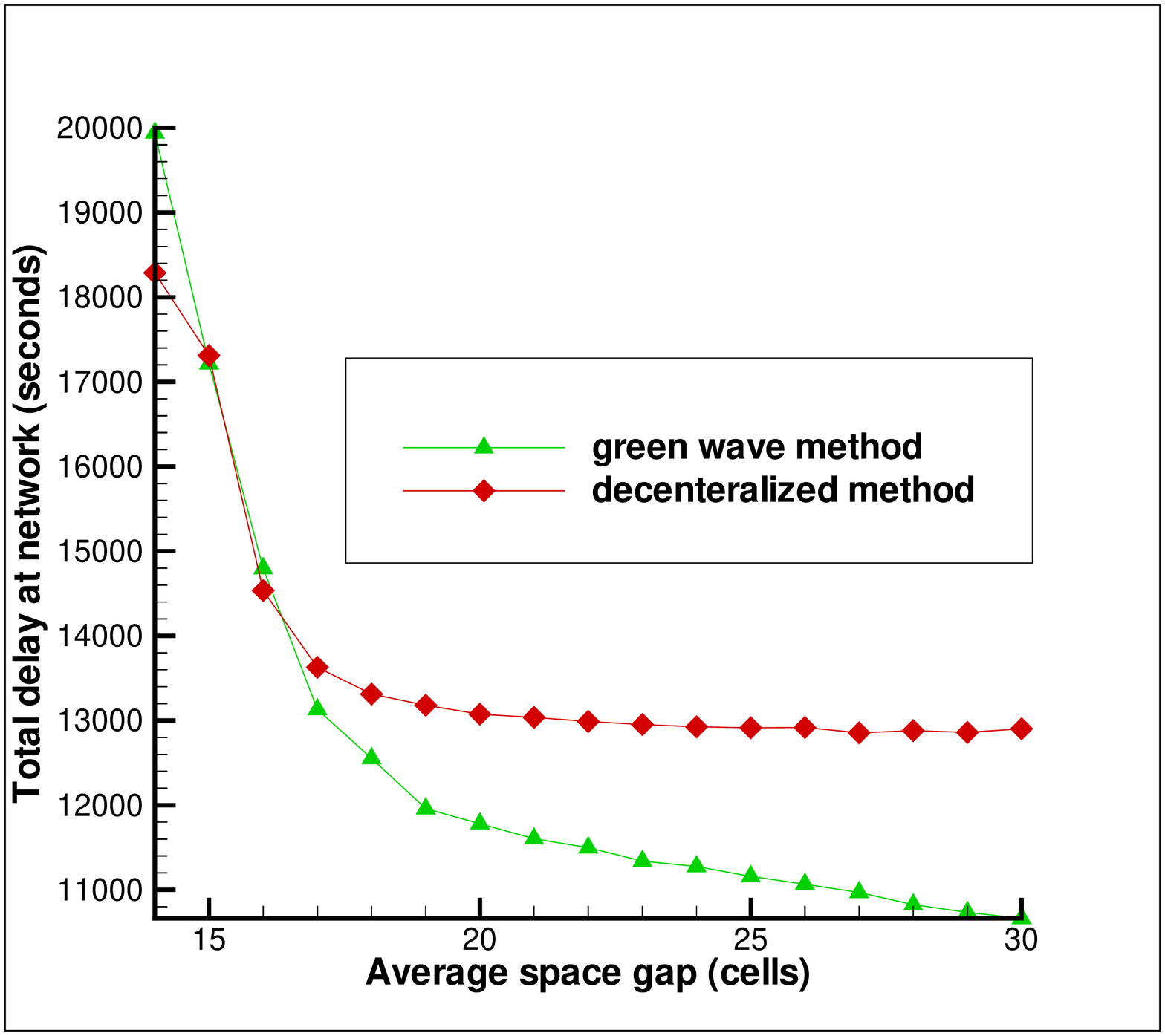}}

\vspace{0.02 cm} {\small{Fig.~7: Overall network delay in terms
of traffic volume compared for both green-wave strategy and
decenteralized scheme. Cut-off lengths are equal to five. Green
time given to the platoons are set to 6 seconds. }}

\end{figure}

According to the above result, the efficiency of the green-wave
strategy is limited to a certain traffic volume. For more
congested situations (corresponding to $\lambda=16$ cells), the
decenteralized scheme acts more optimally than the green-wave
method. The implementation of the green-wave strategy has always
been an argumentative subject. For obvious reasons, the utility of
the green-wave is suppressed when the perpendicular street to the
street in which the green-wave is produced carries a high traffic
volume. This is due to delaying the vehicles of the perpendicular
street so that the platoon of the green-wave street goes through
the green light. Therefore one should be able to determine the
efficiency of the green-wave with respect to the traffic volume in
the perpendicular streets. Simulation reveals that the green-wave
strategy fails to optimise the traffic flow even in our simple
network in the whole range of the traffic volumes. Figure (8)
depicts the overall delay in terms of traffic volumes of the
perpendicular streets A and B (taken equal to each other). The green
wave is produced in street C with constant traffic volume set to
$\lambda_C=14$ cells.

\begin{figure}\label{Fig8}
\epsfxsize=7.2truecm \centerline{\epsfbox{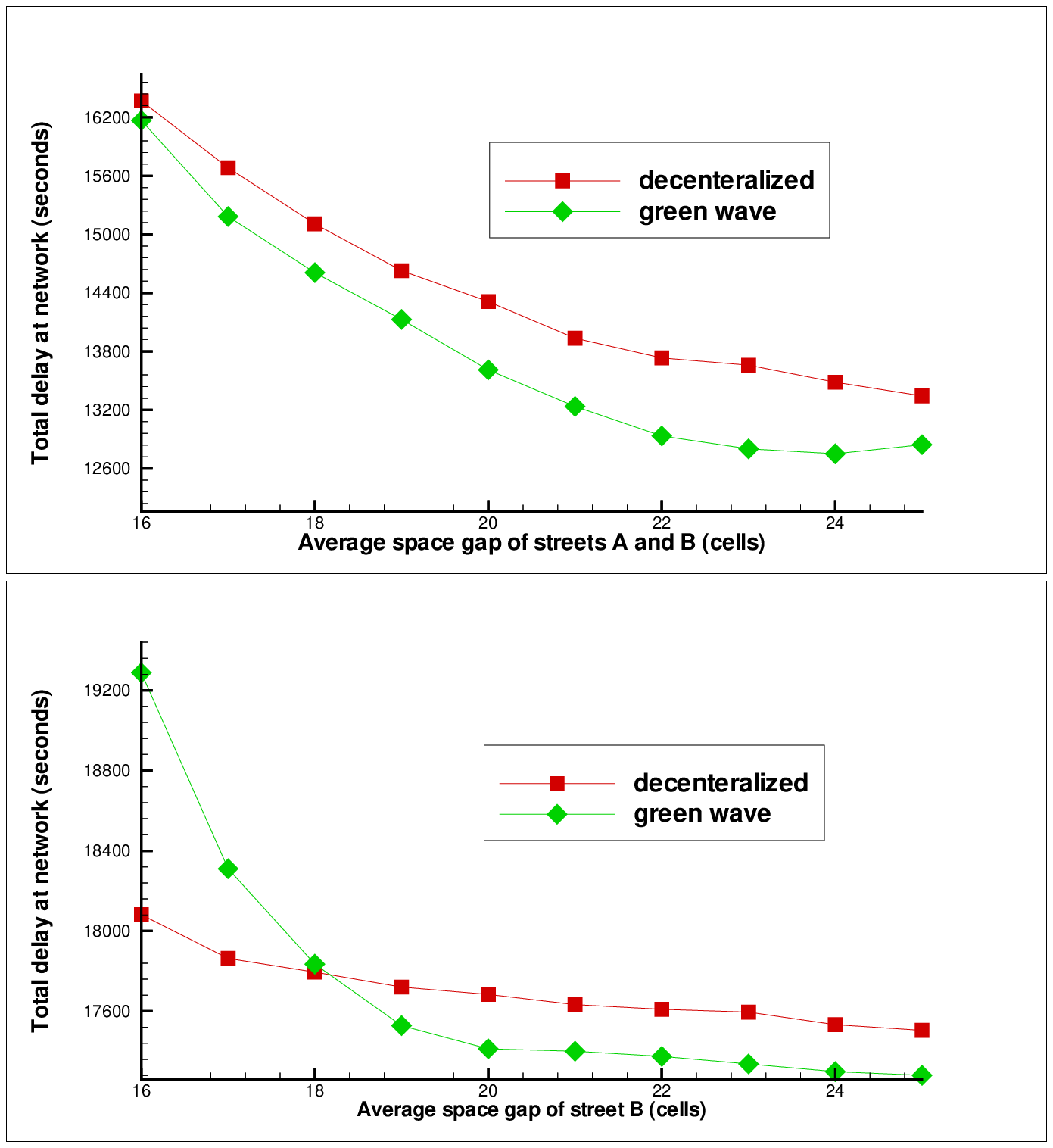}}

\vspace{0.02 cm} {\small{Fig.~8: Overall network delay in terms
of traffic volume in the streets A and B. Green-wave is produced
in street C. Cut-off lengths are equal to five in the up-stream
intersection. }}

\end{figure}

In the top graph of figure (8), green wave efficiency is enhanced
with decreasing of traffic volume in the perpendicular streets.
This is expected since for relatively light traffic volume in
streets A an B, less vehicles are delayed in the red periods when
the platoons are crossing street C. We next investigate the delays
in another asymmetric traffic state in which the traffic volumes
are equal in streets A and C but different to that of street B
(middle graph). We compare the overall delay in green-wave with
decenteralized adaptive scheme by varying $\lambda_B$. One
observes that there is crossover traffic volume for street B
($\lambda_B \sim 18$ cells) below which the optimised traffic flow
through the network is obtained by implementation of
decenteralised scheme. For traffic congestion below this value,
green wave strategy dominated over decenteralised scheme. In order
to gain a better insight to the problem, in the following graphs
we draw the overall delay in terms of two parameters which are the
traffic volumes in the perpendicular streets and the green-time
interval devoted to the platoon approaching to the down-stream
intersection $T_{platoon}^g$.

\begin{figure}\label{Fig9}
\epsfxsize=7.5truecm \centerline{\epsfbox{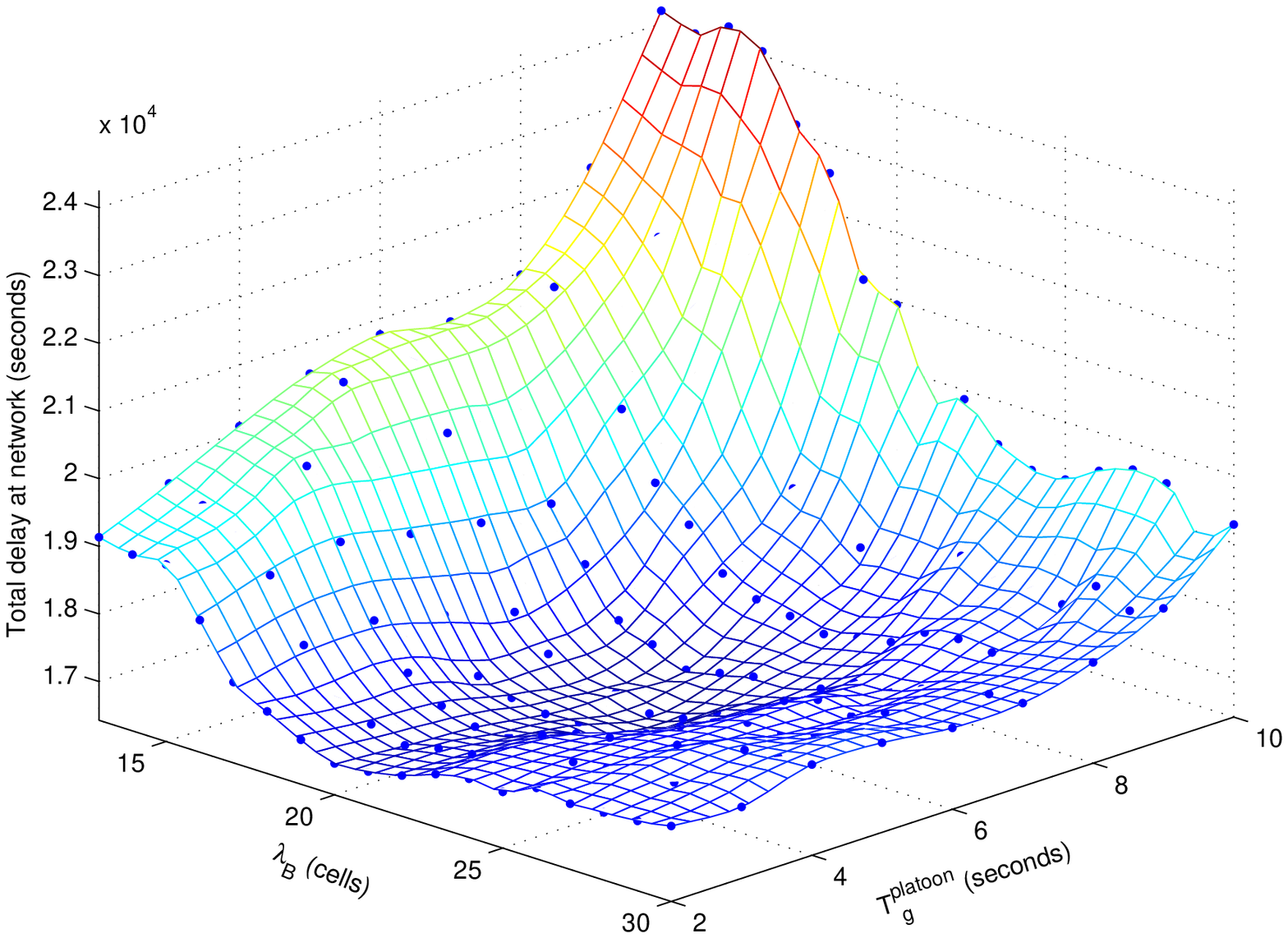}}

\vspace{0.02 cm} {\small{Fig.~9: Overall network delay in terms
of traffic volume in the streets A and B. The green-wave is produced
in street C. Cut-off lengths are equal to five in the up-stream
intersection. Traffic volume is taken as $\lambda=14$ cells street
C. }}

\end{figure}

The above graphs give us the optimal green-time $T_{g}^{platoon}$
devoted to the platoons in terms of the traffic volume on the
perpendicular street. According to the simulation results,
$T_{g}^{platoon}$ not only depends on the platoon length, but also
depends on the traffic volume in the perpendicular streets. This
result could be of practical relevance for real traffic situation.
Our final result (fig. 10) concerns the general comparison of
decenteralized and the green-wave method for the whole traffic
volume range of the perpendicular streets to the green-wave
street. As can be seen, for a wide range of the traffic volume
space, the green-wave method is less efficient with respect to
the decentralised scheme.

\begin{figure}\label{Fig10} \epsfxsize=7.5truecm
\centerline{\epsfbox{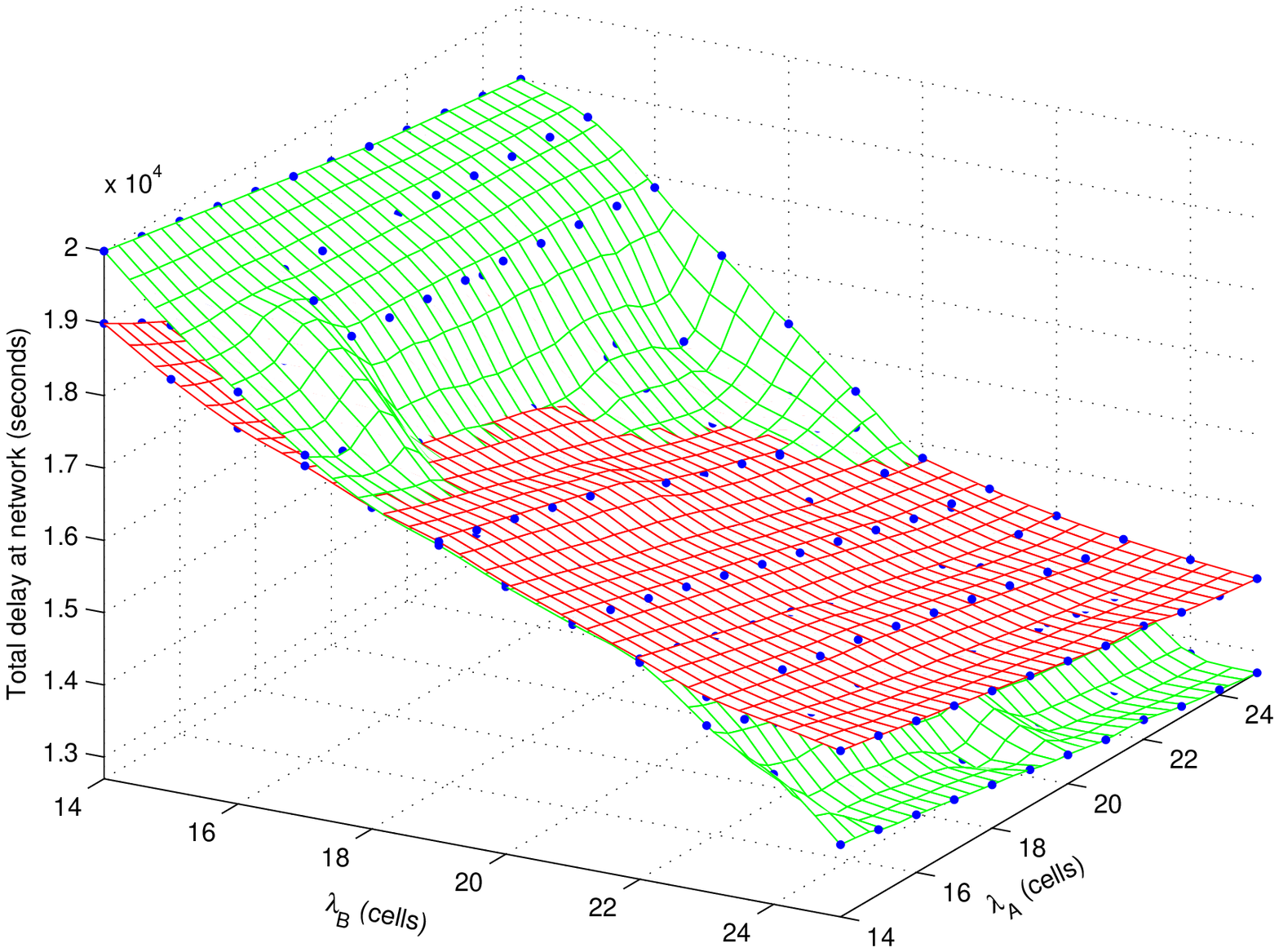}}

\vspace{0.02 cm} {\small{Fig.~10: Overall network delay in terms
of traffic volume in the streets A and B. The green-wave is produced
in street C. Cut-off lengths are equal to five in the up-stream
intersection. Traffic volume is taken as $\lambda=14$ cells street
C. The surface with higher slope corresponds to the 
graan-wave strategy. }}

\end{figure}

\section{Summary and Concluding Remarks}

Recent strides forward, particularly influenced by the contributions from
statistical physics to the subject, have opened new possibilities
for traffic control. In this paper, we have developed and analysed
a prescription for the traffic light signalisation at a small set of linked
intersections. We have proposed an optimising adaptive
decentralized scheme on the basis of {\it minimised total delay }
concept. Borrowing from the traffic engineering literature, we
adopt {\it optimised traffic } as a state in which the total delay
of vehicles is minimum. We have simulated and analysed the {\it
green-wave} method and shown that it is efficient only in a
limited range of traffic volume. We believe that the optimisation
of traffic flow at a small cluster of intersections is a
substantial ingredient towards a global optimisation. Local
clusters of intersections are fundamental operating units of the
sophisticated and correlated urban network and thorough analysis
of them would be advantageous toward the ultimate task of the
global optimisation of the city network. Our simulation results
admit that local adaptive intelligent strategies fail to improve
the traffic state on a global scale. This effect justifies more
explorations on the control of traffic via the coordination of
intersections a subject which is under intensive investigation in
the traffic engineering. We believe the methodologies developed in
the framework of statistical physics to deal with the many body
interactive systems could foster the investigations on city
traffic. Our next objective is the challenge for finding
alternative improved traffic responsive control methods in city
network which are more advantageous over the primitive green-wave
solution. This subject is under our current investigation.

\section{Acknowledgement}

We would like to express our gratitude to Richard W. Sorfleet for reading the manuscript. 
Special thanks are given to Omid Ein-o-din and Mosa Solegh for their useful assistance.

\bibliographystyle{unsrt}

\end{multicols}

\end{document}